\documentclass[twocolumn,american,english,prb,aps,english,superscriptaddress]{revtex4}
\usepackage[T1]{fontenc}
\usepackage[latin9]{inputenc}
\usepackage{color}
\usepackage{amsmath}
\usepackage{amssymb}
\usepackage{cancel}
\usepackage{graphicx}
\usepackage{esint}

\makeatletter
\@ifundefined{textcolor}{}
{%
 \definecolor{BLACK}{gray}{0}
 \definecolor{WHITE}{gray}{1}
 \definecolor{RED}{rgb}{1,0,0}
 \definecolor{GREEN}{rgb}{0,1,0}
 \definecolor{BLUE}{rgb}{0,0,1}
 \definecolor{CYAN}{cmyk}{1,0,0,0}
 \definecolor{MAGENTA}{cmyk}{0,1,0,0}
 \definecolor{YELLOW}{cmyk}{0,0,1,0}
}

\@ifundefined{textcolor}{}{%
 \definecolor{BLACK}{gray}{0}
 \definecolor{WHITE}{gray}{1}
 \definecolor{RED}{rgb}{1,0,0}
 \definecolor{GREEN}{rgb}{0,1,0}
 \definecolor{BLUE}{rgb}{0,0,1}
 \definecolor{CYAN}{cmyk}{1,0,0,0}
 \definecolor{MAGENTA}{cmyk}{0,1,0,0}
 \definecolor{YELLOW}{cmyk}{0,0,1,0}
}

\usepackage{dcolumn}
\usepackage{bm}
\usepackage{enumerate}
\usepackage{babel}

\makeatother

\usepackage{babel}
\begin{document}

\title{Sub-Ohmic to super-Ohmic crossover behavior in nonequilibrium quantum systems with electron-phonon interactions}

\author{Eli Y. Wilner}

\affiliation{School of Physics and Astronomy, The Sackler Faculty of Exact Sciences,
Tel Aviv University,Tel Aviv 69978,Israel}

\author{Haobin Wang}

\affiliation{Department of Chemistry, University of Colorado Denver, Denver, Colorado
80217-3364, USA }

\author{Michael Thoss}

\affiliation{Institute for Theoretical Physics and Interdisciplinary Center for
Molecular Materials, Friedrich-Alexander-Universit{ä}t Erlangen-N{ü}rnberg,
Staudtstr. 7/B2, 91058 Erlangen, Germany}

\author{Eran Rabani}

\affiliation{Department of Chemistry, University of California and Lawrence Berkeley
National Laboratory, Berkeley, California 94720, USA}

\affiliation{The Sackler Center for Computational Molecular and Materials Science,
Tel Aviv University, Tel Aviv, Israel 69978}
\begin{abstract}
The transition from weakly damped coherent motion to localization
in the context of the spin-boson model has been the subject of numerous
studies with distinct behavior depending on the form of the phonon-bath
spectral density, $J\left(\omega\right)\propto\omega^{s}$. Sub-Ohmic
($s<1$) and Ohmic ($s=1$) spectral densities show a clear localization
transition at zero temperature and zero bias, while for super-Ohmic
($s>1$) spectral densities this transition disappears. In this work,
we consider the influence of the phonon-bath spectral density on the
\emph{nonequilibrium} dynamics of a quantum dot with electron-phonon
interactions described by the extended Holstein model. Using the reduced
density matrix formalism combined with the multi-layer multiconfiguration
time-dependent Hartree approach, we investigate the dynamic response,
the time scales for relaxation, as well as the existence of multiple
long-lived solutions as the system-bath coupling changes from the
sub- to the super-Ohmic cases. Bistability is shown to diminish for
increasing powers of $s$ similar to the spin-boson case. However,
the physical mechanism and the dependence on the model parameters
such as the typical bath frequency $\omega_{c}$ and the polaron shift
$\lambda$ are rather distinct. 
\end{abstract}
\maketitle

\section*{Introduction}

Quantum dissipation is an omnipresent phenomenon in diverse physical
systems, ranging from quantum information~\cite{hanson2007spins} and
quantum optics,\cite{Tannoudji1992} to charge transfer and impurity
relaxation,\cite{Marcus1993,Weissbook} superconducting
junctions,\cite{makhlin2001quantum} and more, spanning diverse energy,
length and time scales. Describing the effects of the environment on
the dynamic response of a sub-system requires both the development of
theoretical and computational tools as well as the development of
simplified models necessary to account for the rich system-bath
dynamics and thermodynamic phase behavior. The minimal model required
to capture the essential physics of quantum dissipation involves a
two-level system coupled to a bosonic bath. Perhaps the most studied
version is the well-known spin-boson model,\cite{leggett1987dynamics}
where it is assumed that the two-level system is linearly coupled to a
harmonic bath. The effects of the environment are characterized by the
properties of the bath spectral density, assumed to have a power-law
dependence at low frequencies, $\omega^{s}$
$\left(s\in\mathbb{R}^{+}\right)$, with a cutoff at higher frquencies
determined by a characteristic frequency, $\omega_{c}$. The value of
``$s$'' classifies the nature of the dissipative environment, often
referred to as sub-Ohmic for $0<s<1,$ Ohmic for $s=1$ and super-Ohmic
for $s>1$.

The dynamics, equilibrium and phase behavior of the two-level system
is governed, amongst other factors, by the value of ``$s$''. A notable
transition from coherent to incoherent dynamics in the spin-boson
model is observed as the Kondo parameter, $\eta$ (dimensionless
strength of the system-bath coupling) is increased or when the bath
spectral density changes from super-Ohmic to
sub-Ohmic.\cite{Weissbook,anders2007equilibrium,alvermann2009sparse}
This is followed by a localization transition at high values of $\eta$
in the limit $\omega_{c}\rightarrow\infty$ and low temperature,
$T\rightarrow0$, for the sub-Ohmic and Ohmic cases. Such a transition
disappears
for~\cite{chakravarty1982quantum,bray1982influence,leggett1987dynamics,Weissbook,bulla2003numerical,Serrano-Andres2006}
$s>1$. This rich behavior has been investigated by a variety of
theoretical approaches including
analytical~\cite{Bray82,Chakravarty82,Kehrein96,Spohn89} and
approximate numerical techniques (for an overview, see
Refs.~\onlinecite{leggett1987dynamics,Weissbook}) as well as
numerically exact methods such as numerical renormalization group
(NRG)
techniques,\cite{keil2001real,bulla2003numerical,Anders06,bulla2008numerical}
the multi-configuration` time-dependent Hartree (MCTDH) approach and
its multilayer (ML) extension,
ML-MCTDH,\cite{thoss2001self,wang2008coherent,wang2010coherent} and
path-integral
methods.\cite{Weissbook,Egger94,Makri95b,Muehlbacher03,Kast13}

The dynamics and steady-state properties of dissipative quantum
systems driven away from equilibrium have been the center of more
recent studies, e.g.\ for driven spin-boson-type
systems,\cite{grifoni1998driven} or in the context of inelastic
tunneling in quantum point contacts and molecular
junctions.\cite{klein1973inelastic,adkins1985inelastic,persson1987inelastic,galperin2004inelastic,Persson1998,galperin2004line,Kohler05,repp2005scanning,Haertle08,Cuevas10}
The canonical model in the latter field is given by the extended
Holstein model,\cite{Holstein1959} in which a bridge level (occupied
or empty, hence two levels) is coupled to a bosonic bath, describing
the phonons, and in addition to two fermionic reservoirs representing
the left and right leads.  The latter are held at a different chemical
potentials and thus provide the source to drive the system away from
equilibrium. Most studies focused on steady-state properties utilizing
a variety of techniques to describe, e.g., Franck-Condon
blockade,\cite{koch2005franck,galperin2007inelastic,seldenthuis2008vibrational,Haertle11b,Huetzen12}
negative differential
resistance,\cite{shi2005changes,joachim2005molecular,Galperin05,pati2008origin,Haertle11}
or the existence of multiple long-lived solutions,
i.e.\ bistability.\cite{Bratkovsky2003a,galperin_hysteresis_2005,Mitra2005,Mozyrsky05,Bratkovsky2007,galperin_non-linear_2008,Kosov2011,albrecht_bistability_2012,wilner_bistability_2013}
In addition, novel numerically exact techniques uncovered interesting
transient
behavior.\cite{muhlbacher_real-time_2008,Wang2009,wang_numerically_2011,albrecht_bistability_2012,Simine13}
A promising approach in this regard is the combination of the reduced
density matrix
approach~\cite{cohen_memory_2011,cohen_generalized_2013} and the
multilayer multiconfiguration time-dependent Hartree method in second
quantized representation (ML-MCTDH-SQR),\cite{Wang2009} used to
explore the timescales and the dynamic ``phase diagram'' associated
with the
bistability.\cite{Wang2009,wilner_bistability_2013,wilner2014,wilner_phonon_2014}
The most striking result reported by Wilner \emph{et
  al.~}\cite{wilner2014} for the extended Holstein model with an Ohmic
spectral density is that the bistability persists on timescales
exceeding the phonon-assisted tunneling time along the adiabatic
potential.

At first glance, the localization transition in the spin-boson model
and the bistability in the nonequilibrium extended Holstein model may
result from similar physics. However, these phenomena are quite
different in nature and origin. First, localization in the spin-boson
model is a quantum phase transition strictly at $T=0$, while
bistability is a transient phenomenon persisting over a range of
source-drain bias voltages. Second, the former vanishes in the
adiabatic limit of a slow bath ($\omega_{c}\rightarrow0$) while the
latter thrives in this limit. Finally, in the Ohmic case, localization due to a degenerate ground state and the corresponding dependence of the steady state on the initial state occurs only for the symmetric spin-boson model, while bistability in the extended Holstein model spans a wide range of asymmetries

In this work, we explore the dependence of bistability on the nature
of the phonon spectral density using the combined ML-MCTDH-SQR and
reduced density matrix approach. We cover both the the sub-Ohmic to
super-Ohmic limits. While localization and bistability show opposite
behavior with respect to $\omega_{c}$, we find similarties with
respect to the boson spectral density. Specifically, as $s$ is
increased above $1$ the bistability diminishes. This transition,
however, is not as sharp as the localization transition.

\section*{Model Hamiltonian and Spectral Densities}

To describe the effect of different forms of dissipation on
nonequilibrium transport in a quantum system with electron-phonon
interaction, we consider a generic model Hamiltonian describing, e.g.,
a quantum dot or a molecular junction:
\begin{equation}
H=H_{S}+H_{B}+V_{SB}\label{eq:Hamiltonian}.
\end{equation}
Here, $H_{S}=\varepsilon_{d}d^{\dagger}d$ is the system Hamiltonian
representing the electronic degrees of freedom of the quantum dot with
creation/annihilation fermionic operators $d^{\dagger}$/$d$ and energy
$\varepsilon_{d}$. For simplicity, we assume that the quantum dot is
represented by a single level. The bath Hamiltonian,
$H_{B}=H_{\ell}+H_{{\rm ph}}$, is given as a sum of electron (lead)
and phonon baths where

\begin{equation}
H_{\ell}=\sum_{k\in L,R}\varepsilon_{k}a_{k}^{\dagger}a_{k}
\end{equation}
represents the noninteracting leads Hamiltonian with fermionic
creation/annihilation operators $a_{k}^{\dagger}$/$a_{k}$, and

\begin{equation}
H_{ph}=\sum_{j}\hbar\omega_{j}\left(b_{j}^{\dagger}b_{j}+\frac{1}{2}\right)
\end{equation}
represents the phonon bath with creation/annihilation bosonic
operators $b_{j}^{\dagger}$/$b_{j}$ for phonon mode $\alpha$
with energy $\hbar\omega_{j}$. The coupling between the system
and the baths is given by
\begin{equation}
V_{SB}=\sum_{k\in L,R}\left(t_{k}da_{k}^{\dagger}+t_{k}^{*}a_{k}d^{\dagger}\right)+d^{\dagger}d\sum_{j}M_{j}\left(b_{j}^{\dagger}+b_{j}\right)\label{eq:V}
\end{equation}
where $t_{k}$ is the hopping term between the system and the leads and
$M_{\alpha}$is the strength of the electron-phonon couplings to mode
$\alpha$. The former is determined from the relation
\begin{equation}
\Gamma_{L,R}(\varepsilon)=2\pi\sum_{k\in L,R}|t_{k}|^{2}\delta(\varepsilon-\varepsilon_{k}),\label{eq:Gamma}
\end{equation}
with
$\Gamma_{L,R}\left(\varepsilon\right)=\frac{a^{2}}{b^{2}}\sqrt{4b^{2}-(\varepsilon-\mu_{L,R})^{2}}$
used to mimic a tight-binding chain and $\mu_{L,R}$ is the chemical
potential of the left/right lead, respectively. We adopt the same
parameters for $\Gamma_{L,R}\left(\varepsilon\right)$ used in our
recent studies,\cite{wilner_bistability_2013,wilner2014} namely,
$a=0.2\mbox{eV}$ and $b=1\mbox{eV}$. For this choice,
$\Gamma=0.16\mbox{eV}$ is the maximum value of
$\Gamma_{R}\left(\varepsilon\right)+\Gamma_{L}\left(\varepsilon\right)$.
The electron-phonon couplings, $M_{\alpha}$, are determined from the
relation:
\begin{equation}
J\left(\omega\right)=\pi\sum_{j}M_{j}^{2}\delta\left(\hbar\omega-\hbar\omega_{j}\right)\label{eq:J(w)}
\end{equation}
\begin{figure}[t]
\includegraphics[width=8cm]{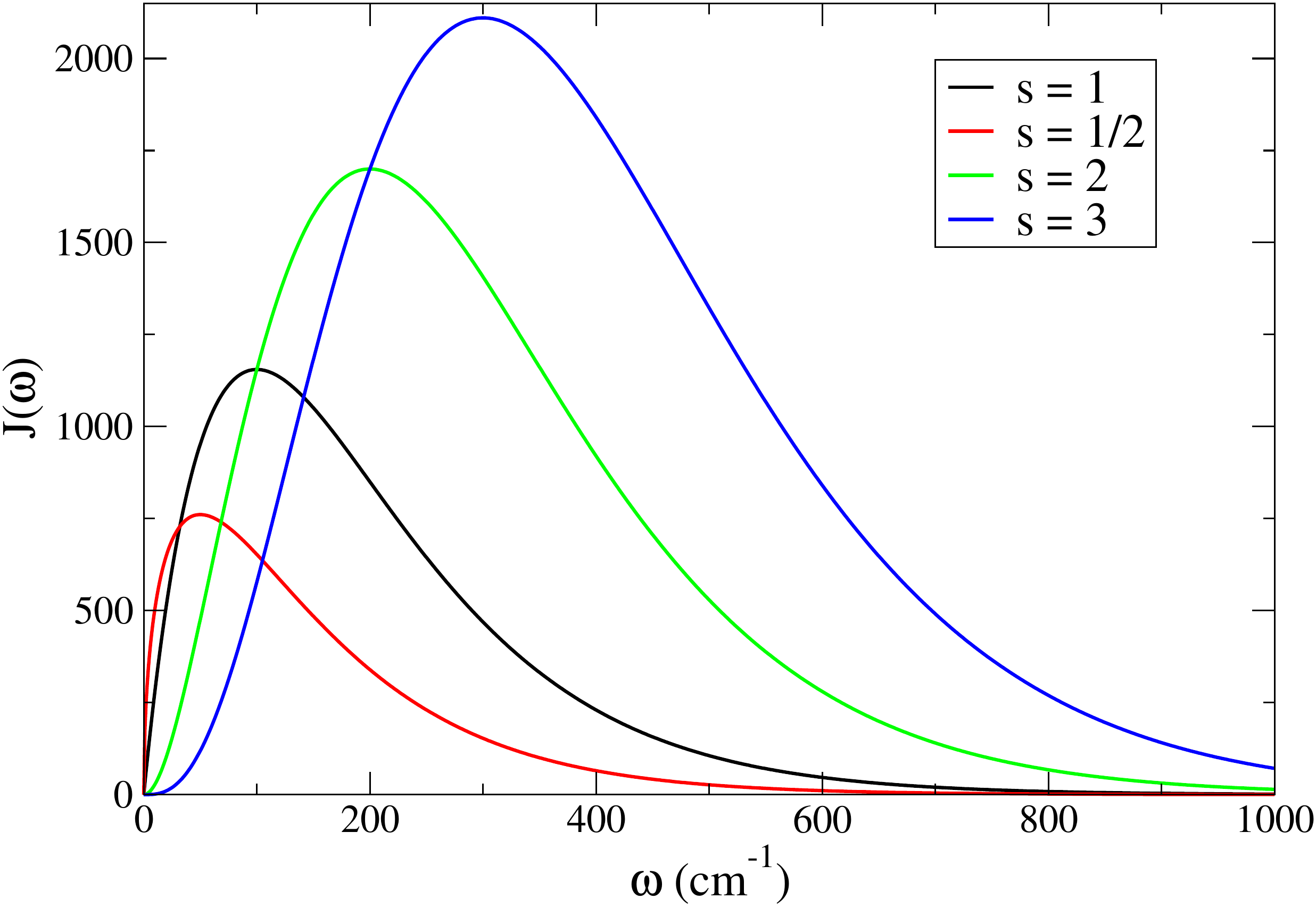}\protect\protect\caption{\label{fig:spectral}
  The phonon spectral density,
  $J\left(\omega\right)=\frac{\lambda\pi\omega^{s}}{\Gamma\left(s\right)\omega_{c}^{s}}e^{-\frac{\omega}{\omega_{c}}}$,
  for various values of $s$ for
  $\omega_{c}=100\mbox{cm\ensuremath{^{-1}}and
    \ensuremath{\lambda=1000\mbox{cm\ensuremath{^{-1}.}}}}$}
\end{figure}
where we follow the notation of Caldeira and Leggett~\cite{Caldeira81}
for the phonon spectral density:
\begin{equation}
J\left(\omega\right)=\frac{\pi\hbar}{2}\eta\left(\omega^{s}/\omega_{c}^{s-1}\right)e^{-\frac{\omega}{\omega_{c}}}.
\end{equation}
In the above equation, the dimensionless Kondo parameter,
$\eta=\frac{2\lambda}{\hbar\omega_{c}\Gamma\left(s\right)}$, determines the
overall strength of the electron-phonon couplings, $\omega_{c}$ is the
characteristic phonon bath frequency and $\Gamma\left(s\right)$ is the
Euler Gamma function. For future reference, we introduce an additional
parameter
$\lambda=\sum_{j}\frac{M_{j}^{2}}{\hbar\omega_{j}}=\frac{1}{\pi}\int\frac{d\omega}{\omega}J(\omega)$,
which is known as the reorganization energy (or polaron
shift). Fig.~\ref{fig:spectral} shows the results for
$J\left(\omega\right)$ for different values of $s$.

The model introduced above and variants thereof have been widely used
to study nonequilibrium charge transport in nanostructures, such as,
for example, semiconductor quantum dots,\cite{Kubala10} carbon
nanotubes~\cite{Leturcq09} or molecular
junctions.\cite{Cizek04,Mitra2004,Koch05,galperin_hysteresis_2005,Galperin07,Benesch08}
In the latter case, the phonons may include, in addition to the
phonons of the contacts, the vibrational degrees of freedom of the
molecule.  In all previous studies, however, the spectral density was
limited to the Ohmic case, for which $s=1$. The sub-Ohmic ($s<1)$ and
super-Ohmic ($s>1$) limits which play an important role in the related
spin-boson model,\cite{Weissbook} have not been studied in transport
junctions.

\section*{Reduced Density Matrix}

The dynamics generated by the above Hamiltonian are rich and rather
complicated to solve. Numerically exact techniques include real time
path
integrations~\cite{muhlbacher_real-time_2008,weiss_iterative_2008,werner_diagrammatic_2009,eckel_comparative_2010,werner_weak-coupling_2010,Segal10,gull10_bold_monte_carlo,Huetzen12,Simine13}
and ML-MCTDH
approach.\cite{thoss2001self,wang2008coherent,wang2010coherent} Both
are limited to relatively short times and cannot describe the dynamics
on all relevant timescales. Recently, we have proposed to combine a
numerically exact impurity solver with a reduced density matrix
formalism.\cite{cohen_memory_2011,Cohen2013kondo,wilner_bistability_2013,cohen_generalized_2013,wilner2014}
Application to the above model for the Ohmic spectral density
uncovered a fascinating behavior with rich dynamics on multiple
timescales and bistability persisting on timescale longer than the
phonon--assisted tunneling
times.\cite{wilner_bistability_2013,wilner2014,wilner_phonon_2014}
Here, we adopt this approach to study the influence of different forms
of the phonon spectral density on the dynamic response and on the
bistability. For completeness, we briefly review the formalism.

The basic quantity of interest is the reduced density matrix,
$\sigma\left(t\right)$, which is derived from the full density matrix,
$\rho\left(t\right)$, by the application of the projection operator
$P=\rho_{B}Tr_{B}$.  Here, the index $B$ refers to the bath degrees of
freedom or the ``irrelevant'' part of the full
Hamiltonian. $\sigma\left(t\right)=Tr_{B}\rho\left(t\right)$ obeys a
generalized quantum master equation, given
by:\foreignlanguage{american}{\cite{zhang_nonequilibrium_2006}}

\selectlanguage{american}%
\begin{equation}
i\hbar\frac{\partial}{\partial t}\sigma\left(t\right)=\mathcal{L}_{S}\sigma\left(t\right)+\vartheta\left(t\right)-\frac{i}{\hbar}\int_{0}^{t}d\tau\kappa\left(\tau\right)\sigma\left(t-\tau\right).\label{eq:sigma(t)}
\end{equation}
\foreignlanguage{english}{where $\mathcal{L}_{S}=[H_{S},\cdots]$
is the system's Liouvillian, 
\begin{equation}
\vartheta\left(t\right)=Tr_{B}\left\{ \mathcal{L}_{V}e^{-\frac{i}{\hbar}Q\mathcal{L}t}Q\rho\left(0\right)\right\} \label{eq:theta(t)}
\end{equation}
depends on the choice of initial conditions and vanishes for an
uncorrelated initial state (which is the case considered below),
\textit{i.e.}, when $\rho(0)=\sigma(0)\otimes\rho_{B}(0)$, where
$\sigma(0)$ and $\rho_{B}(0)$ are the system and bath initial density
matrices, respectively, and $\mathcal{L}_{v}=[V_{SB},\cdots]$. We
consider two initial conditions for $\sigma\left(0\right)$, an
occupied and unoccupied dot. We assume a non-correlated initial state
for$\rho_{B}(0)$ }

\selectlanguage{english}%
\begin{equation}
\rho_{B}\left(0\right)=\rho_{\mbox{ph}}\left(0\right)\otimes\rho_{\ell}^{L}(0)\otimes\rho_{\ell}^{R}(0),\label{eq:rho0}
\end{equation}
where ($\beta=\frac{1}{k_{B}T}$ is the inverse temperature) 
\begin{equation}
\rho_{\ell}^{L/R}(0)=\exp\left[-\beta\left(\sum_{k\in L/R}\left(\varepsilon_{k}-\mu_{L/R}\right)a_{k}^{\dagger}a_{k}\right)\right],\label{eq:rholeads}
\end{equation}
is the initial density matrix for the leads, and

\begin{align}
\rho_{\mbox{ph}}\left(0\right) & =\exp\left[-\beta\left\{ \sum_{\alpha}\hbar\omega_{j}\left(b_{j}^{\dagger}b_{j}+\frac{1}{2}\right)\right.\right.\nonumber \\
 & \left.\left.+\sum_{j}\delta_{j}M_{j}\left(b_{j}^{\dagger}+b_{j}\right)\right\} \right]\label{eq:rhoph}
\end{align}
represents the initial density matrix of the phonon bath. We also
consider two different initial conditions for the phonons, one where
$\delta_{j}=0$ in Eq.~\eqref{eq:rhoph} corresponding to phonons
initially equilibrated with an unoccupied dot, and another where
$\delta_{j}=1$ corresponding to phonons equilibrated to an
occupied dot. More details can be found in
Ref.~\onlinecite{wilner_bistability_2013}.

\begin{figure*}
\includegraphics[width=8cm]{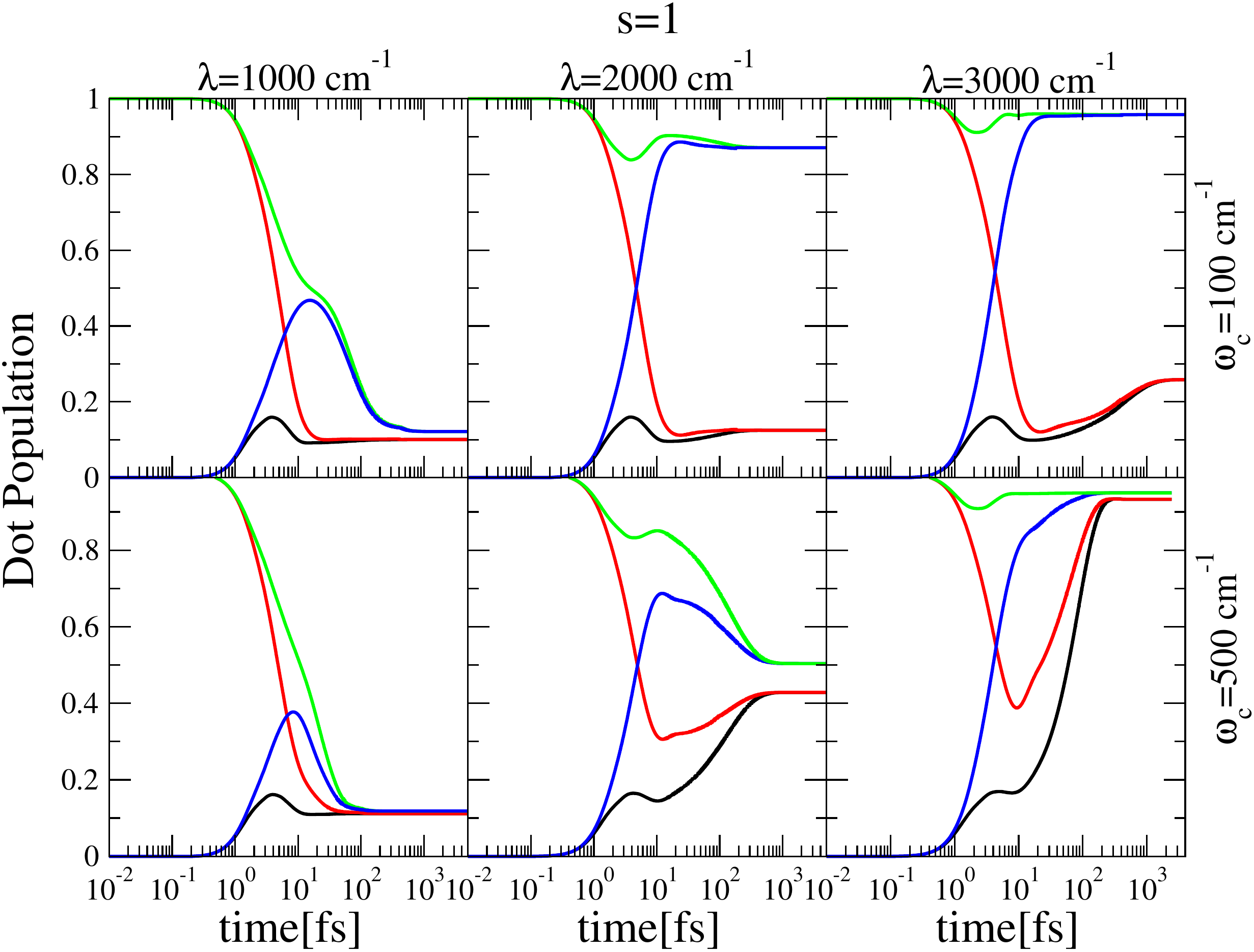}\includegraphics[width=8cm]{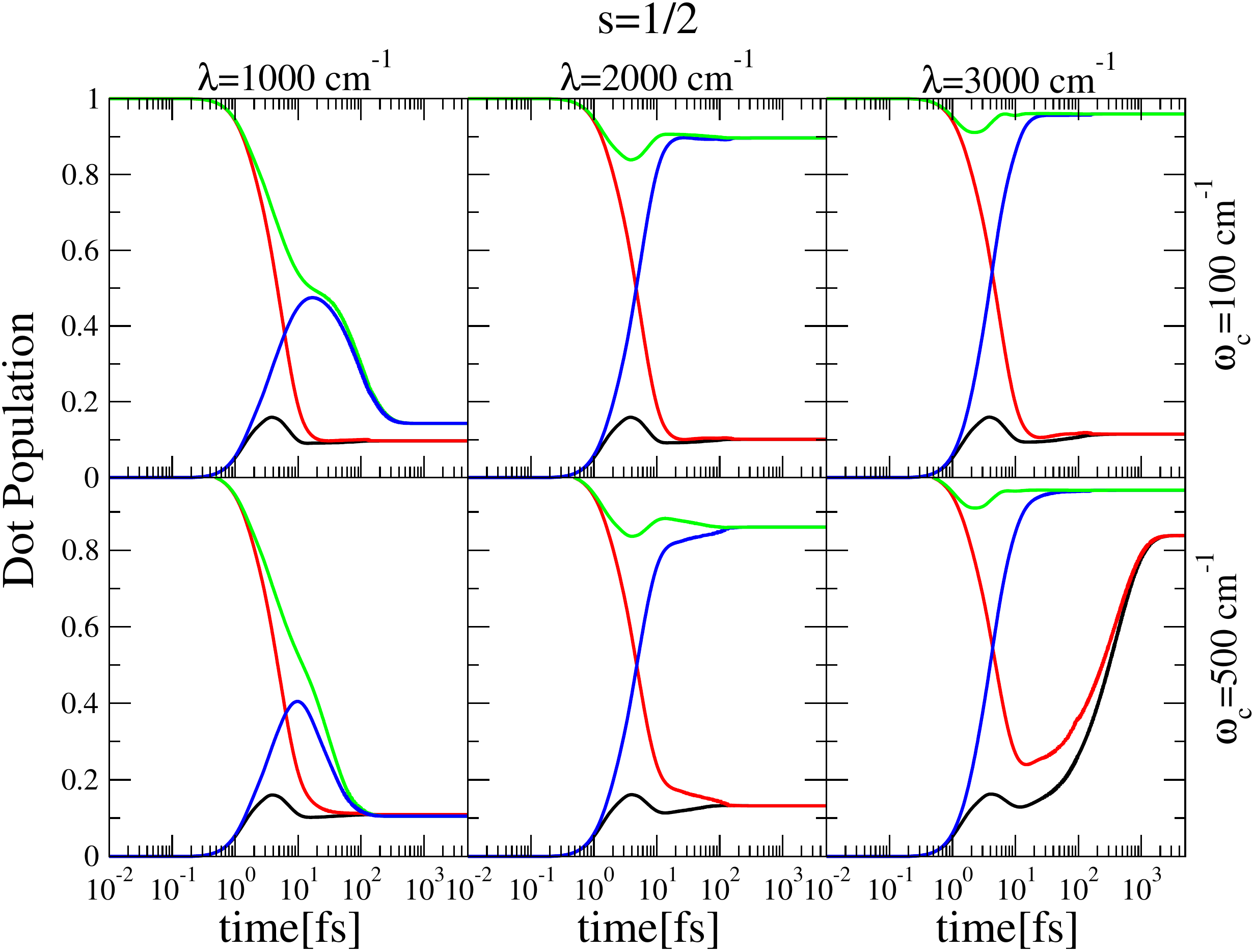}\protect\protect\caption{\label{fig:sub-ohmic}
  Transient dynamics of the average quantum dot population for the
  Ohmic (left set of panels) and sub-Ohmic case (right set of
  panels). The results are shown for all $4$ initial conditions: Black
  and red curves correspond to unoccupied / occupied dot at phonon
  initial condition $\delta_{j}=0$, whereas blue and green curves
  correspond to unoccupied / occupied at$\delta_{j}=1$,
  respectively.  In all results the cutoff time used to generated the
  memory kernel is $t_{c}=100\mbox{fs}.$}
\end{figure*}

The calculations of the memory kernel in Eq.~\eqref{eq:sigma(t)},
$\kappa\left(t\right)$, is the tricky part. Formally it is given by
\begin{equation}
\kappa\left(t\right)=Tr_{B}\left\{ \mathcal{L}_{V}e^{-\frac{i}{\hbar}Q\mathcal{L}t}Q\mathcal{L}\rho_{B}\right\} \label{eq:memory-kernel}
\end{equation}
where $Q=1-P$, $P=\rho_{B}(0)Tr_{B}\{\cdots\}$ and
$\mathcal{L}=[H,\cdots]$.  A more suitable form for the memory is
given in terms of a Volterra equation of the second type, removing the
complexity of the projected dynamics of Eq.~(\ref{eq:memory-kernel}):
\begin{equation}
\kappa\left(t\right)=i\hbar\dot{\Phi}\left(t\right)-\Phi\left(t\right)\mathcal{L}_{S}+\frac{i}{\hbar}\int_{0}^{t}d\tau\Phi\left(t-\tau\right)\kappa\left(\tau\right)\label{eq:volterra}
\end{equation}
with (since the operator $\mathcal{L}_{V}$ appearing in the equation
for $\Phi(t)$ and the full Hamiltonian conserve the total particle
number, only the diagonal matrix elements $\Phi(t)$ need to be
computed):
\begin{equation}
\Phi_{nn,mm}(t)=\frac{2}{\hbar}Tr_{B}\left\{ \rho_{B}\left\langle m\right|\sum_{k}t_{k}d(t)a_{k}^{\dagger}(t)\left|m\right\rangle \right\} ..\label{eq:phi(t)}
\end{equation}
Here, $|m\rangle$ denotes the electronic state of the quantum dot,
where $m$ can take the values $1$ or $0$, corresponding to an occupied
or an unoccupied dot, respectively. Previously, we have shown that
$\Phi_{nn,mm}(t)$ can be expressed in terms of the sum of the left
($I_{m}^{L}(t)$) and right ($I_{m}^{R}(t)$) currents:

\begin{equation}
e\Phi_{nn,mm}(t)=I_{m}^{L}(t)+I_{m}^{R}(t),\label{eq:current-1}
\end{equation}
where 
\begin{equation}
I_{m}^{L,R}(t)=-\frac{2e}{\hbar}\Im\sum_{k\in L,R}t_{k}\langle m|d(t)a_{k}^{\dagger}(t)|m\rangle,\label{eq:current}
\end{equation}
is the left/right current for an initial occupied ($m=1$) or empty
($m=0$) dot, and $e$ is the electron charge.

While the calculation of the memory kernel requires a solution of the
time-dependent left and right currents, it typically decays on
timescales much faster than the density matrix
itself,\cite{cohen_memory_2011,wilner_bistability_2013} and thus is
amenable to a numerically exact impurity solvers. 
For this purpose we adopt the ML-MCTDH
approach~\cite{thoss2001self,wang2008coherent,wang2010coherent} and
calculate the memory up to a cutoff time $t_{c}$, where $t_{c}$ is
large enough such that the results for the reduced density matrix do
not change for large cutoff times. When suitable, i.e., for small
electron-phonon couplings,\cite{wilner2014} we often use the
nonequilibrium Green's function formalism within the self-consistent
Born approximation to generate the memory kernel for $t<t_{c}$.

\section*{Sub-Ohmic and Ohmic cases}

\begin{figure*}
\includegraphics[width=8cm]{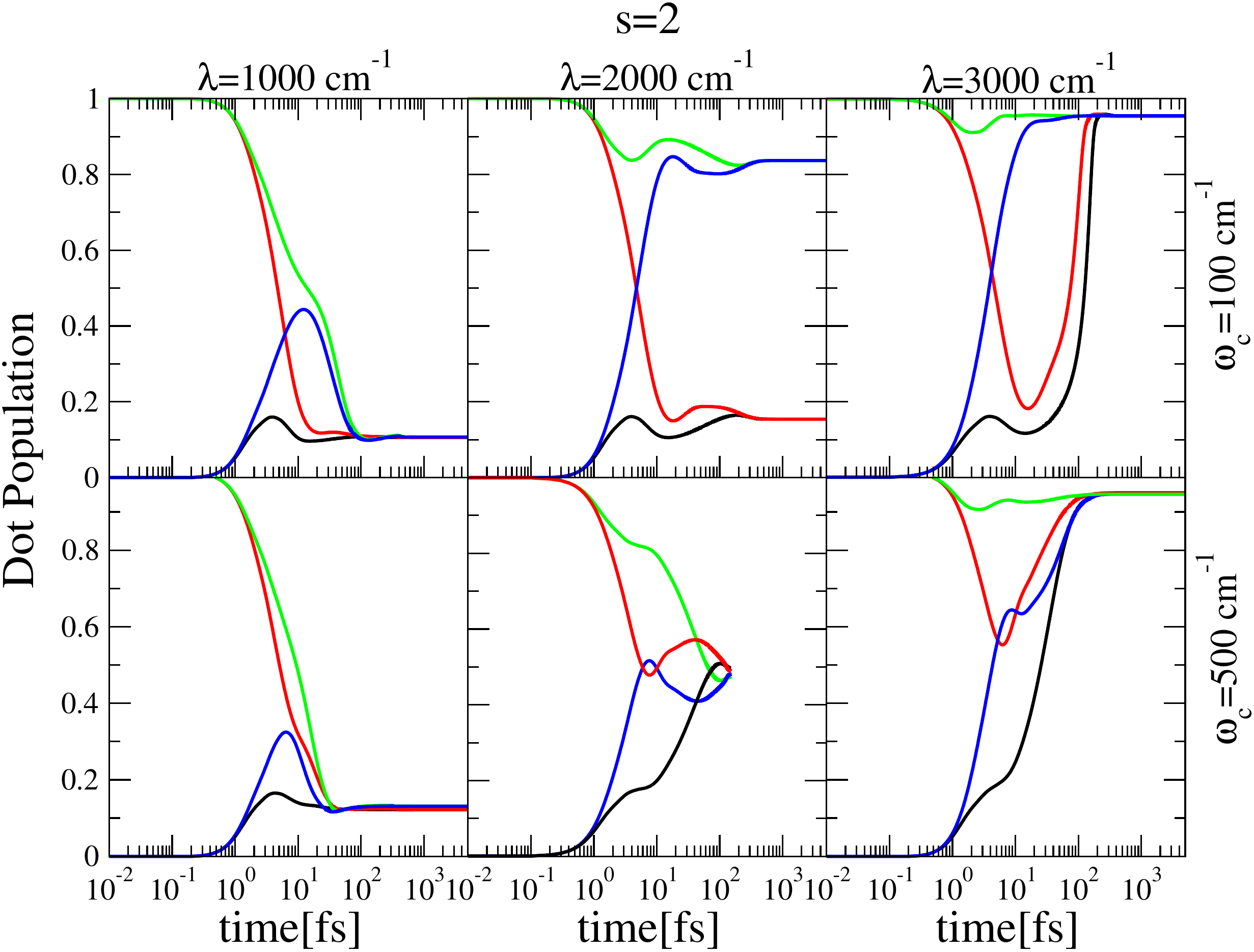}\includegraphics[width=8cm]{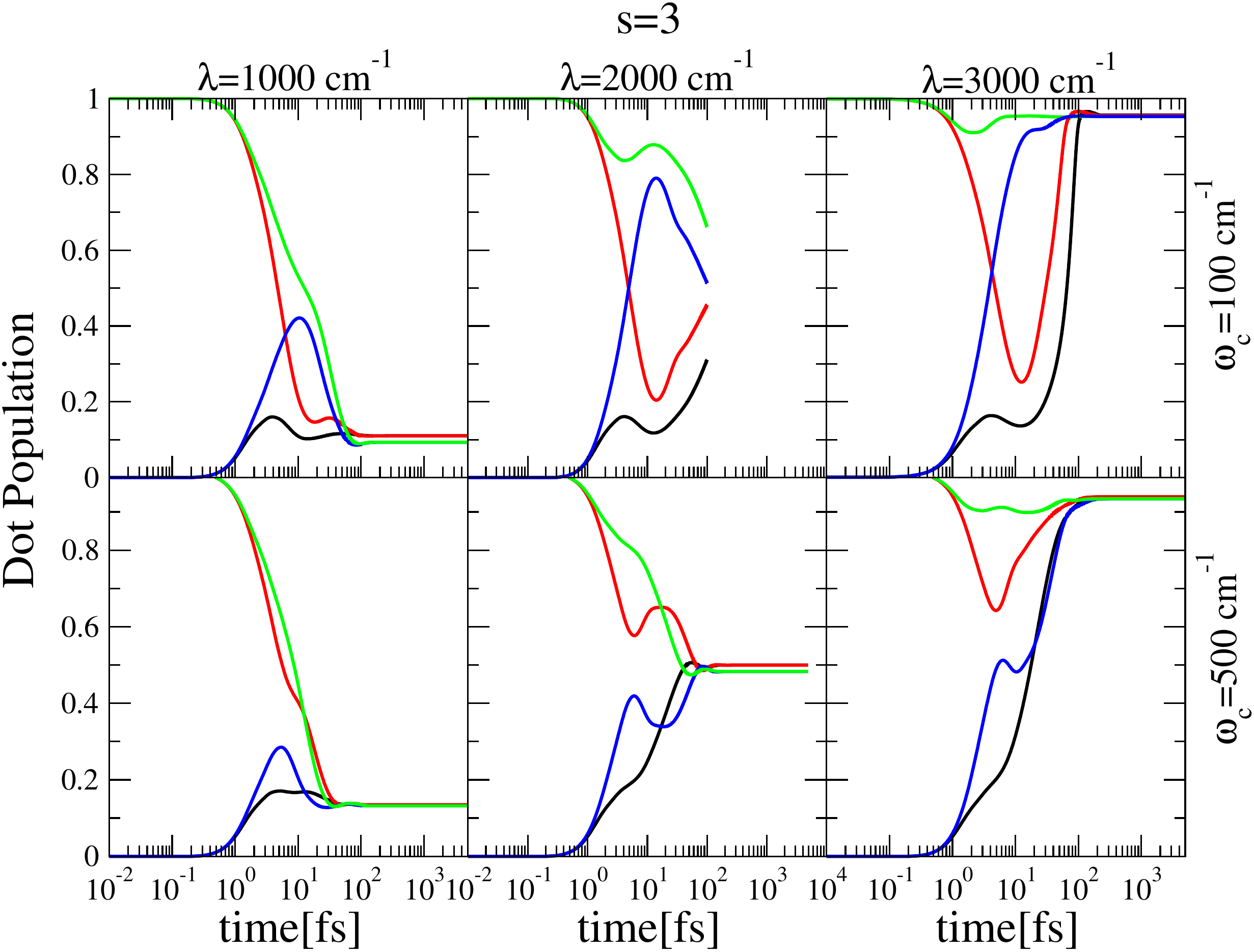}\protect\protect\caption{\label{fig:super-ohmic}
  Transient dynamics of the average quantum dot population for the
  super-Ohmic case. The results are shown for all $4$ initial
  conditions: Black and red curves correspond to unoccupied / occupied
  dot at phonon initial condition $\delta_{j}=0$, whereas blue
  and green curves correspond to unoccupied / occupied
  at$\delta_{j}=1$, respectively. In all results the cutoff time
  used to generated the memory kernel is $t_{c}=100\mbox{fs}.$
  Converging the results $\omega_{c}=500\mbox{cm}^{-1}$,
  $\lambda=2000\mbox{cm}^{-1}$for $s=2$ and
  $\omega_{c}=100\mbox{cm}^{-1}$, $\lambda=2000\mbox{cm}^{-1}$for
  $s=3$ is difficult, and thus we only show the direct ML-MCTDH-SQR to
  $t=t_{c}$.}
\end{figure*}

In Fig.~\ref{fig:sub-ohmic} we show the populations dynamics for the
Ohmic $\left(s=1\right)$ and sub-Ohmic $\left(s=\frac{1}{2}\right)$
cases for various values of the reorganization parameter $\lambda$ and
the phonon frequency $\omega_{c}$. Throughout this work we fix the
electronic parameters: Dot energy $\varepsilon_{d}=0.25\mbox{eV}$,
applied bias-voltage $\mu_{L}=-\mu_{R}=0.05\mbox{eV}$ and zero
temperature ($k_{{\rm B}}T=0$). The results shown for the Ohmic case
(left set of panels in Fig.~\ref{fig:sub-ohmic}) summarize our
previous findings.\cite{wilner_bistability_2013,wilner2014} In short,
the decay of the population is characterized by three distinct
timescales. At short to intermediate times, the population dynamics
are governed by the system-lead hybridization $\hbar/\Gamma$(as
clearly seen for $\delta_{j}=0$,
$\lambda=1000\mbox{cm\ensuremath{^{-1}}}$) or by $\omega_{c}^{-1}$ (as
clearly seen for $\delta_{j}=1$,
$\lambda=1000\mbox{cm\ensuremath{^{-1}}}$). In addition to the short
and intermediate timescales associated with the separate electronic
and phononic degrees of freedom, the electron-phonon coupling
introduces longer timescales related to the tunneling between the two
charge states, as clearly evident for increasing values of
$\lambda$. On timescale longer than this tunneling time, the
population is characterized by two distinct, long-lived, solutions for
large values of $\lambda$.  This ``bistability'' vanishes as the
system become less adiabatic (increasing $\omega_{c}$) or for small
polaron shifts (small $\lambda$).  Readers interested in a
comprehensive discussion of the Ohmic case are encouraged to consult
Ref.~\onlinecite{wilner2014}.

The situation is similar for the sub-Ohmic case (right set of panels
in Fig.~\ref{fig:sub-ohmic}). All three timescales are clearly
observed even for $s=\frac{1}{2}$. For
$\lambda=1000\mbox{cm}^{-1}<\varepsilon_{d}$, the more stable solution
is associated with the non-shifted configuration ($\delta_{j}=0$)
and the system relaxes to a single long-lived state regardless of the
choice of initial condition ($\delta_{j}=0$ or $1$). The dynamics
at short and intermediate times are governed by $\hbar/\Gamma$ for the
non-shifted phonon bath ($\delta_{j}=0$), since the dot energy
$\varepsilon_{d}=0.25\mbox{eV}$ is outside the conduction window
($\Delta\mu=0.1\mbox{eV}$) and thus the system remains in the
uncharged state. This is not the case for the shifted bath initial
preparation ($\delta_{j}=1$). Since the effective dot energy
($\tilde{\varepsilon}_{d}=\varepsilon_{d}-2\lambda$,
Ref.~\onlinecite{wilner2014}) is within the conduction window, we
observe a rapid uncharging decay associated with $\hbar/\Gamma$
followed by a slower relaxation to the stable uncharged configuration
along the generalized bath mode, characterized by $1/\omega_{c}$
timescale.

As $\lambda$ increases above $\varepsilon_{d}$, the long-time behavior
of the population depends on the choice of the initial phonon
preparation, leading to a bistability (two distinct long lived
solutions). The bistability is rather sensitive to the characteristic
phonon frequency, and can lead to two steady states solutions in the
adiabatic limit, as
$\omega_{c}\rightarrow0$.\cite{galperin_non-linear_2008} As
$\omega_{c}$ increases the bistability gradually disappears. Comparing
the sub-Ohmic case to the Ohmic case, the bistability persists over a
wider range of frequencies and polaron shifts for $s<1$. This can be
explained qualitatively by looking at the spectral density function
shown in Fig.~\ref{fig:spectral}. In the sub-Ohmic case, the bath
spectrum is shifted to the lower frequency end as compared with the
Ohmic case that has the same characteristic frequency
$\omega_{c}$. Thus, a comparison of the dynamics generated for the
Ohmic case at a particular values of $\omega_{c}$ should be done with
the dynamics generated for the sub-Ohmic case ($s=1/2$) for a higher
$\omega_{c}$.

To summarize the sub-Ohmic case, we find that the behavior is similar
to the Ohmic case. The major difference is observed for the long time
relaxation. The sub-Ohmic case shows a pronounced bistability at
larger values of $\omega_{c}$ and a wider range of polaron shifts
compared to the Ohmic case, resulting from the increase in the density
of low frequency modes as $s$ is decreased below 1.

\section*{Super-Ohmic case}

\begin{figure*}
\includegraphics[width=8cm]{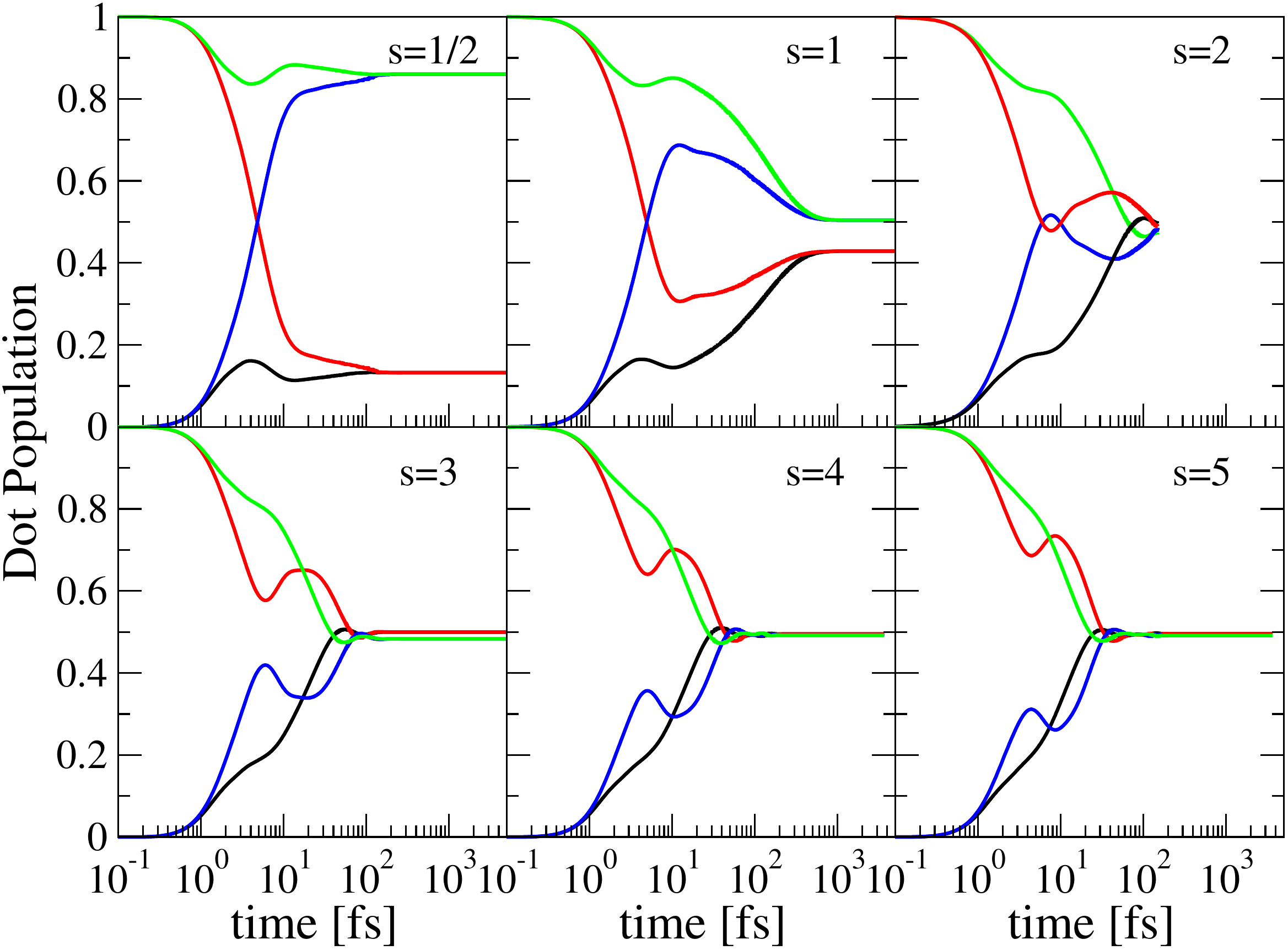}\includegraphics[width=8cm]{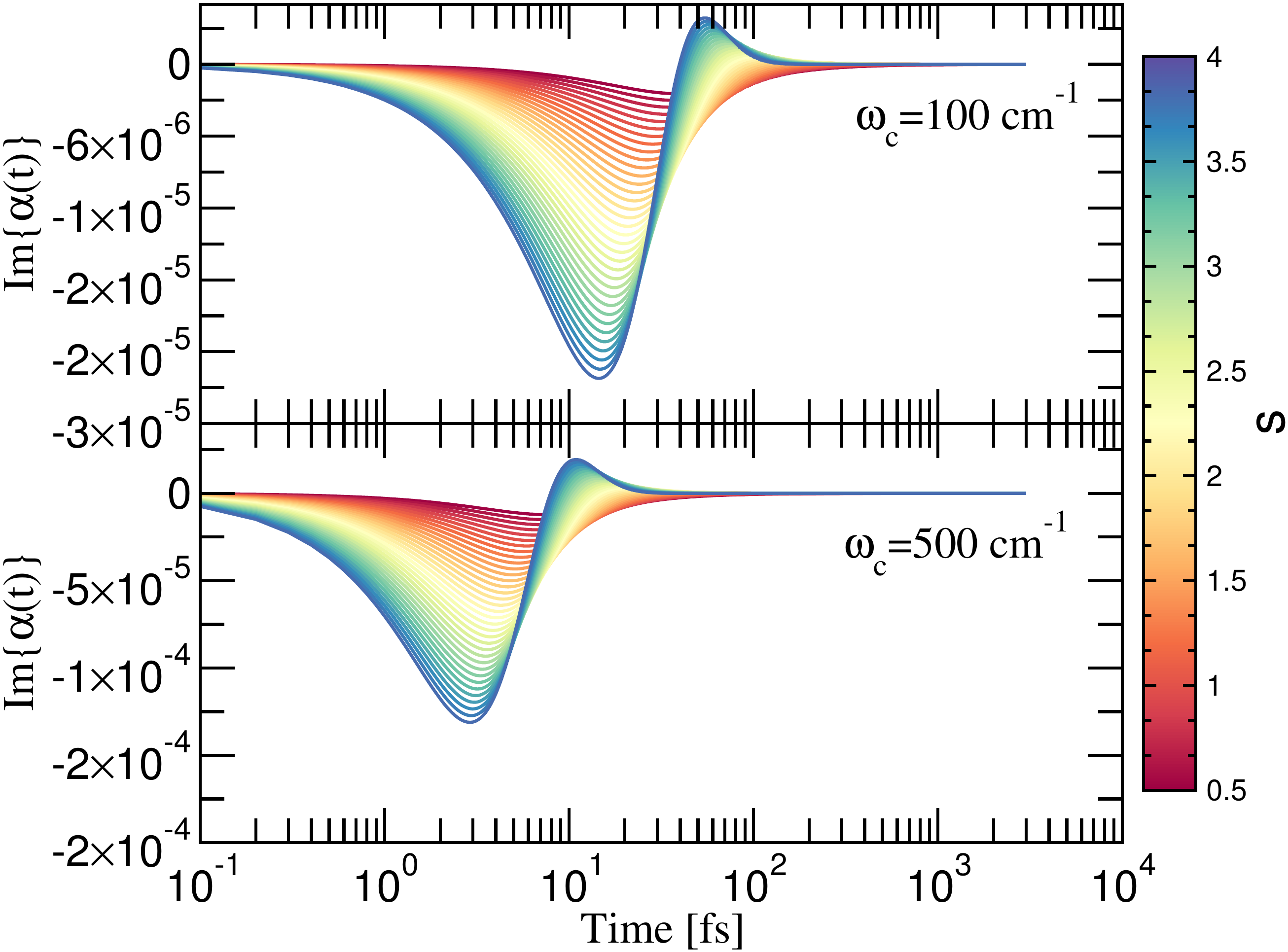}\protect\protect\caption{\label{fig:super-Ohmic-2}
  Left: Dot population in the super-Ohmic case for
  $\varepsilon_{d}=0.25\mbox{eV}$, $\lambda=2000\mbox{cm}^{-1}$ and
  $\omega_{c}=500\mbox{cm}^{-1}$. The results are shown for all $4$
  initial conditions: Black and red curves correspond to unoccupied /
  occupied dot at phonon initial condition $\delta_{j}=0$,
  whereas blue and green curves correspond to unoccupied / occupied
  at$\delta_{j}=1$, respectively. Right: $\Im\left\{
  \alpha\left(t\right)\right\} $ for different value of $s$ for
  $\omega_{c}=100\mbox{cm}^{-1}$ (upper panel) and
  $\omega_{c}=500\mbox{cm}^{-1}$(lower panel). }
\end{figure*}

The trends observed going from the sub-Ohmic case to the Ohmic case
continue smoothly as $s$ is increased above $1$. In
Fig.~\ref{fig:super-ohmic} we plots the transient population dynamics
for the super-Ohmic case ($s=2$ and $s=3$) for the same set of
frequencies and polaron shifts shown in Fig.~\ref{fig:sub-ohmic}. In
two cases, we are unable to converge the results based on the memory
formalism with the given input and thus, show the direct calculations
based on the ML-MCTDH-SQR approach generated for
$t<t_{c}=100\mbox{fs}$. Similar to sub-Ohmic case, for
$\lambda=1000\mbox{cm}^{-1}$ the population dynamics at short and
intermediate times is governed by $\hbar/\Gamma$ for the non-shifted
phonon bath ($\delta_{j}=0$) and by the typical phonon frequency
for the shifted bath initial preparation ($\delta_{j}=1$).  The
major effect associated with changing $s$ is the change in the
characteristic phonon frequency, given by $s\omega_{c}$.  As $s$
increases above $1$, the characteristic frequency increases leading to
a rapid decay of the population at intermediate times, evident in
Figs.~\ref{fig:sub-ohmic} and \ref{fig:super-ohmic}.  Concerning the
long time behavior, we find that the tunneling dynamics are washed out
as $s$ increases above $1$ and the bistability is limited to a
narrower range of polaron shifts, which eventually disappears for
$s\ge3$ for the range of frequencies studied.

Comparing the behavior of the bistability to the localization
transition in the spin-boson model, it is clear that the origin of the
two phenomena is quite different.  While in the spin-boson model
localization is a quantum phase transition, observed strictly at zero
temperature and zero bias, bistability is a dynamical phenomenon
observed in the time-domain. Furthermore, while localization vanishes
for $s>1$, the bistability transitions smoothly across the Ohmic
case. Moreover, the two phenomena differ and show opposing behavior
with respect to the dependence on $\omega_{c}$ and to a large extent
with respect to the polaron shift $\lambda$. For the latter, the
bistability is limited to a certain window of $\lambda$, a window
which smoothly narrows down as $s$ increases. Localization in the
spin-boson model persists above a certain value of $\lambda$.

\begin{figure}[h]
\includegraphics[width=8cm]{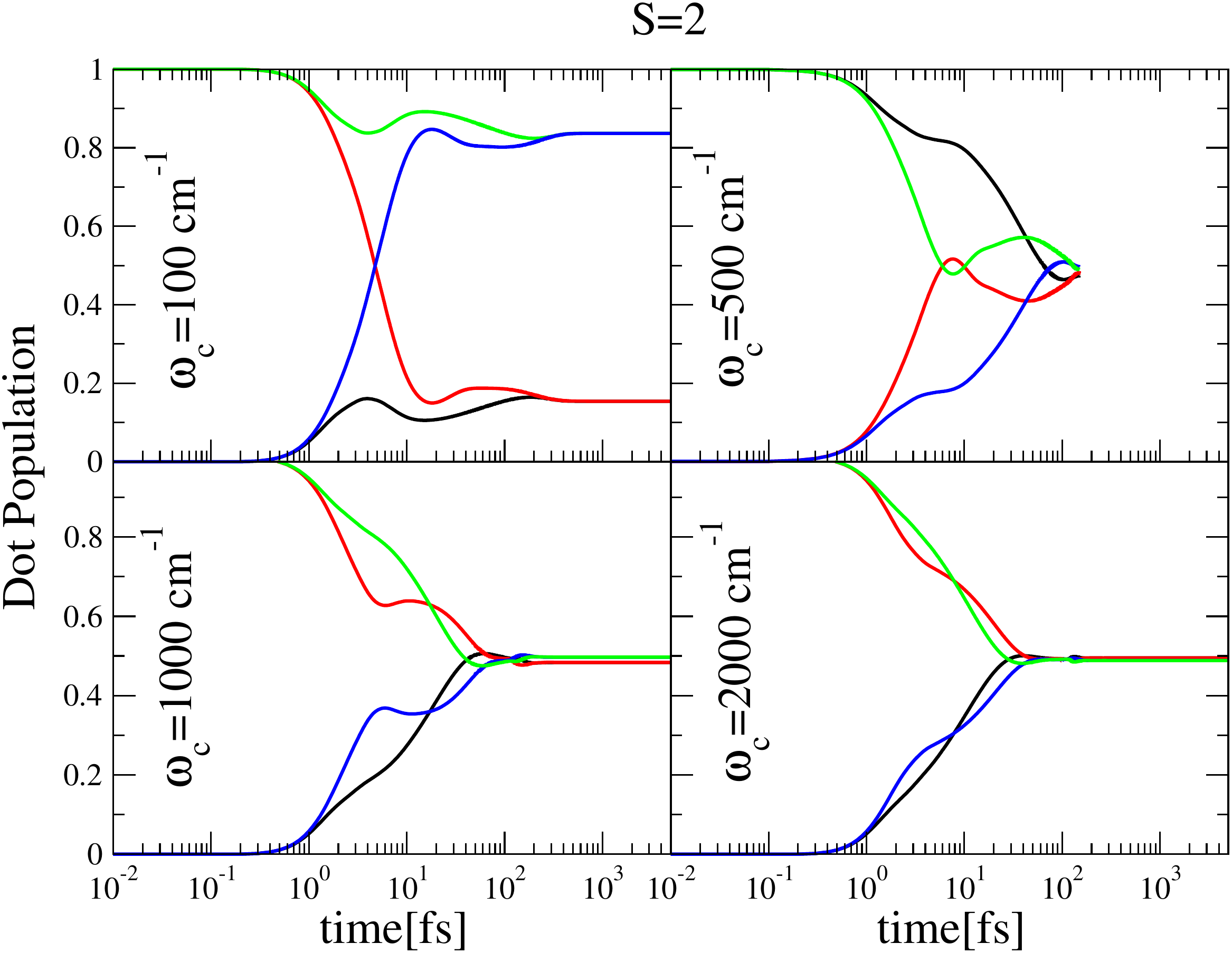}\protect\protect\caption{\label{fig:w-c}
 Coherent to incoherent transition in the super-Ohmic case $s=2$. Shown are results of the dot population for different characteristic frequencies $\omega_c$ of the phonon bath as indicated in the plot and for four different initial condition as indicated by the different colors. The color code is the same as in Fig.\ref{fig:super-Ohmic-2}}
\end{figure}

A notable feature of the super-Ohmic case is the pronounced
oscillations in the population compared to the over-damped Ohmic and
sub-Ohmic behavior, as shown in the left panels of
Fig.~\ref{fig:super-Ohmic-2}.  This is similar to the coherent
dynamics observed for the super-Ohmic case in the spin-boson
model. The period of oscillations increases as the characteristic
frequency of the boson bath decreases, which makes the convergence of
the memory formalism difficult for $\omega_{c}$ values between 100 and
500 $\mbox{cm}^{-1}$ due to the relatively long transient dynamics. As
$\omega_{c}$ increases, the coherence is eventually quenched because
of the more efficient energy exchange between the electron and phonon
degrees of freedom, as shown in Fig.~\ref{fig:w-c}.

To understand this behavior, we propose to examine the phonon bath
autocorrelation function, given by (at $T=0$): 
\begin{equation}
\label{eq:alpha-t}
\alpha\left(t\right)=\frac{1}{\hbar\pi}\int_{0}^{\infty}d\omega\,J\left(\omega\right)\exp\left(-i\omega
t\right).
\end{equation}
Using the spectral density of Eq.~\eqref{eq:J(w)}, we obtain an exact
solution to $\alpha\left(t\right)$ of the
form:
\begin{align}
\alpha\left(t\right) & =\frac{\eta\omega_{c}^{2}}{2}\frac{\Gamma\left(1+s\right)}{\left(1+t^{2}\omega_{c}^{2}\right)^{\frac{s+1}{2}}}\nonumber \\
 & \times\exp\left[-i\left(1+s\right)\arctan\left(\omega_{c}t\right)\right].
 \label{alpha}
\end{align}
In the right panels in Fig.~\ref{fig:super-Ohmic-2} we plot the
imaginary part of $\alpha\left(t\right)$ for two values of the
frequency and for different values of $s$. $\alpha\left(t\right)$
shows a transition from a smooth function of time at small values of
$s<1$ to an oscillatory function at larger values of $s>1$. The origin
of this oscillatory behavior is in the term
$\exp\left[i\left(1+s\right)\arctan\left(\omega_{c}t\right)\right]$
and is correlated with the under-damped dynamics of the population
shown in the left panels of Fig.~\ref{fig:super-Ohmic-2}.

A more quantitative picture of the oscillatory behavior emerges by
considering the dynamics of the reaction mode $\left\langle
Q\left(t\right)\right\rangle
=\frac{\sum_{\alpha}M_{\alpha}\left\langle
  b_{j}^{\dagger}\left(t\right)+b_{j}\left(t\right)\right\rangle
}{\sqrt{\sum_{\alpha}2M_{\alpha}^{2}}}$.\cite{wilner_phonon_2014}
Despite the fact that the dynamics of the phonons have been traced out
by considering the reduced density matrix of the system alone,
$\left\langle Q\left(t\right)\right\rangle $ for the non-shifted case
($\delta_{j}=0$) can be inferred using its equation of motion,
resulting in~\cite{wilner_phonon_2014}
\begin{eqnarray}
\left\langle Q\left(t\right)\right\rangle  & \mbox{=} & 
Q_1(t) + Q_2(t)
\label{eq:b(t)}
\end{eqnarray}
with
\begin{eqnarray}
Q_1(t) & \mbox{=} &  n_{\infty} \sqrt{\frac{2}{\alpha(0)}}
\Im\int_{0}^{t}\alpha\left(\tau\right)\mbox{d} \ensuremath{\tau}\label{Q1} \\
 Q_2(t) & \mbox{=} & \sqrt{\frac{2}{\alpha(0)}}\Im\int_{0}^{t}\delta n\left(\tau\right)\alpha\left(t-\tau\right)\mbox{d}\tau
 \label{Q2},
\end{eqnarray}
where $n_{\infty}$ is the steady state dot population and $\delta
n\left(t\right)=\left\langle
d^{\dagger}\left(t\right)d\left(t\right)\right\rangle -n_{\infty}$.  A
similar expression not given here can be derived for the shifted case
($\delta_{j}=1$) (see appendix).  The first term, $Q_1$, on the right hand side
of Eq.~(\ref{eq:b(t)}) describes the relaxation of the reaction mode
due to the coupling to the phonon bath in the absence of coupling to
the leads. It can be calculated explicitly using Eq.~(\ref{alpha}),
resulting in
\begin{eqnarray}
Q_1(t) & \mbox{=} & 
n_{\infty}\sqrt{\frac{2\lambda}{s\hbar\omega_c}}
\left( \frac{\cos(s\arctan\left(\omega_{c}t\right))}{\left(1+t^{2}\omega_{c}^{2}\right)^{\frac{s}{2}}}
-1 \right)
\label{Q1_expl},
\end{eqnarray}
where the relations $2\alpha(0) = \eta\omega_c^2 \Gamma(1+s)$,
$\int_0^{\infty} \alpha(\tau)d\tau = -i\lambda/\hbar$ and $\eta = 2\lambda /
(\hbar\omega_c \Gamma(s))$ were used.  The second term, $Q_2(t)$, includes
the influence of the coupling to the leads on the dynamics of the
reaction mode. The long-time limit of the reaction mode is determined
by the first term in Eq.~(\ref{eq:b(t)}), i.e.\
\begin{equation}
\lim_{t\to \infty} \langle Q(t)\rangle = \lim_{t\to \infty}Q_1(t) =-n_{\infty}\sqrt{\frac{2\lambda}{\hbar s\omega_c}} 
\end{equation}

\begin{figure}[t]
\includegraphics[width=8cm]{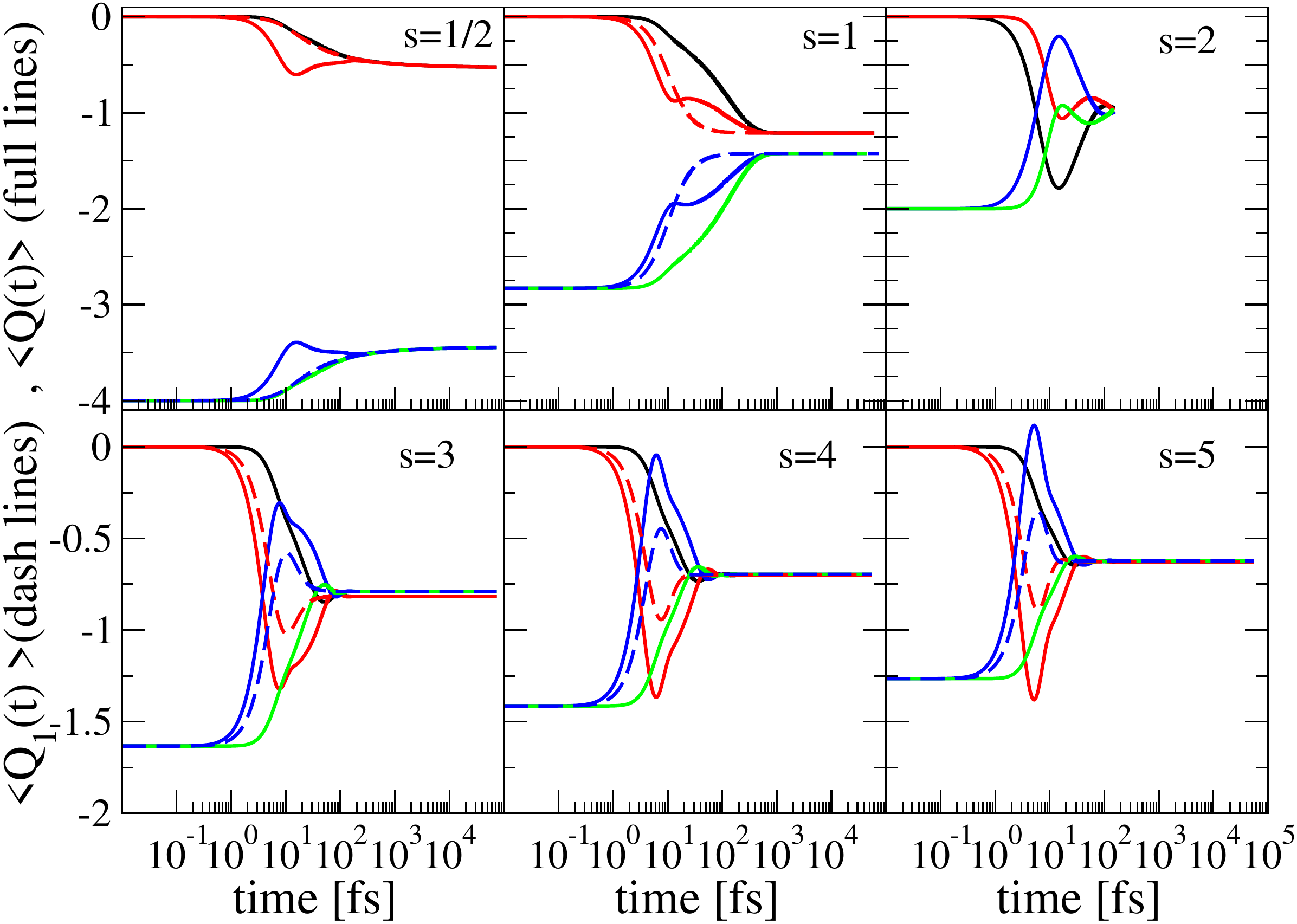}\protect\protect\caption{\label{fig:reaction-coordiante}The
  reaction coordinate $\left\langle Q\left(t\right)\right\rangle $ for
  several values of $s$ and for $\varepsilon_{d}=0.25\mbox{eV}$,
  $\lambda=2000\mbox{cm}^{-1}$ and $\omega_{c}=500\mbox{cm}^{-1}$ and
  temperature $T=0$. Black - unoccupied $n_{d}\left(0\right)=0$ and
  $\delta_{j}=0$; \textcolor{red}{Red} - occupied
  $n_{d}\left(0\right)=1$ and $\delta_{j}=0$;
  \textcolor{blue}{Blue} - unoccupied $n_{d}\left(0\right)=0$ and
  $\delta_{j}=1$; \textcolor{green}{Green} - occupied
  $n_{d}\left(0\right)=1$ and $\delta_{j}=1$. In addition, the
  dashed lines depict $Q_1(t)$, the part of $\left\langle
  Q\left(t\right)\right\rangle $ which describes the relaxation of the
  reaction mode due to the coupling to the phonon bath in the absence
  of coupling to the leads (see Eq.~(\ref{Q1})).}
\end{figure}

In Fig.~\ref{fig:reaction-coordiante} we plot the dynamics of the
reaction mode for the same parameters used to generate the data in
left set of panels Fig.~\ref{fig:super-Ohmic-2} for both the sub-Ohmic
and super-Ohmic cases. Note that in both figures, we are unable to
converge the results for $s=2$ beyond the cutoff time
$t_{c}=100\mbox{fs}$ (see the discussion above).  For $s\le1$, the
reaction coordinate $\left\langle Q\left(t\right)\right\rangle $
decays monotonically to its steady state value if it is initially in
equilibrium with the electronic state of the dot (full black and green
lines). The timescale governing this decay can either be purely
phononic, arising from the term $Q_1(t)$ with an algebraic relaxation
characteristics given by $\sim
(1+t^{2}\omega_{c}^{2})^{-\frac{s}{2}}$,
or associated with the relaxation of the electronic population $\delta
n\left(t\right)$ arising from the term $Q_2(t)$.  For $s=\frac{1}{2}$
the decay of $\left\langle Q\left(t\right)\right\rangle $ is
significantly slower than that of $\delta n\left(t\right)$ and thus is
determined by the slower phononic dynamics, while for $s=1$ there is
no clear time scale separation. In case of a preparation, where the
phonon degrees of freedom are initially not in equilibrium with the
electronic state of the dot (full red and blue lines), a more complex
transient dynamics is seen, which involves electronic and phononic
contributions and time scales.

In both cases ($s=\frac{1}{2}$ and $1$), the dot population assumes
for longer times two distinct values depending on the initial
conditions for the phonon bath, i.e.\ exhibits bistability, as clearly
evident in Fig.~\ref{fig:super-Ohmic-2}.  The bistability in the dot
populations leads to two solutions for $\left\langle
Q\left(t\right)\right\rangle $. In the long-time limit, the difference
between the solutions for $\left\langle Q\left(t\right)\right\rangle
$, obtained for shifted and unshifted initial preparation, is related
to the corresponding difference of the dot population $\Delta n_{d}$
by~\cite{wilner_phonon_2014,note1}
\begin{equation}
\Delta Q=-\frac{\lambda}{\hbar}\sqrt{\frac{2}{\alpha\left(0\right)}}\Delta n_{d} 
= -\sqrt{\frac{2 \lambda}{s\hbar \omega_c}}\Delta n_{d} .
\end{equation}
This relation between $\Delta n_{d}$ and $\Delta Q$ is only valid in
the steady state and can therefore be used as a consistency check if
the steady state has been reached. For the results in
Fig.~\ref{fig:reaction-coordiante} it is fulfilled.

As $s$ is increased above $1$ the picture changes qualitatively.
First, the bistability gradually disappears as discussed in details
above. This implies that $\left\langle
Q\left(t\rightarrow\infty\right)\right\rangle $ assumes a unique value
regardless of the initial preparation of the phonons. Second, as
clearly evident in
Fig.~\ref{fig:reaction-coordiante}, $\left\langle
Q\left(t\right)\right\rangle $ shows a pronounced oscillatory behavior
correlated with the oscillations observed in the dot population shown
in Fig.~\ref{fig:super-Ohmic-2}. Most interestingly, for the unshifted
initial phonon preparation ($\delta_{j}=0$), the oscillation in
$\left\langle Q\left(t\right)\right\rangle $ lead to transient values
for the reaction coordinate that are associated with the shifted
position, and vice versa. This ``exchange'' of positions is also
reflected, to a smaller extent, in the dot population shown in the
corresponding panels of Fig.~\ref{fig:super-Ohmic-2}.  For larger
values of $s$ ($s > 1$), the bare phonon relaxation time scale $\sim
(1+t^{2}\omega_{c}^{2})^{-\frac{s}{2}}$ is significantly faster than
the electronic dynamics $\delta n\left(t\right)$, and thus the latter
dominates the dynamics of the reaction mode $\left\langle
Q(t)\right\rangle $ for longer times.

\section*{Conclusions}

In this work we have studied the role of the spectral density of the
phonon bath on the relaxation dynamics and bistability in the
nonequilibrium extended Holstein model. To this end, sub-Ohmic, Ohmic
and super-Ohmic spectral densities were considered. The results show a
physically rich behavior, including a transition from incoherent to
oscillatory dynamics and a disappearance of bistability signatures
upon increase of the power $s$ of the spectral density.
Some features observed for the sub- and super-Ohmic cases can be
rationalized qualitatively by rescaling the characteristic phonon
frequency, $\omega_{c}\rightarrow s\omega_{c}$.
For example, the bistability which is exemplified in adiabatic limit
$\omega_{c}\rightarrow0$, persists over a larger range of frequencies
for sub-Ohmic spectral density while the opposite is true for the
super-Ohmic case. However, the appearance of a slow algebraic decay
for $\left\langle Q\left(t\right)\right\rangle $ in the sub-Ohmic case
as well as the oscillatory behavior of $\left\langle
Q\left(t\right)\right\rangle $ in the super-Ohmic case cannot be
explained by simple scaling arguments, but require a more elaborate
analysis in terms of the bath correlation function,
$\alpha\left(t\right)$ and are distinct from the Ohmic limit.
\begin{acknowledgments}
EYW is grateful to The Center for Nanoscience and Nanotechnology at
Tel Aviv University for a doctoral fellowship. HW acknowledges the
support from the National Science Foundation CHE-11500285. MT
acknowledges support by the German Research Council (DFG) and the
German-Israeli Foundation for Scientific Research and Development
(GIF). This work used resources of the National Energy Research
Scientific Computing Center, which is supported by the Office of
Science of the U.S. Department of Energy under Contract
No. DE-AC02-05CH11231, and the Leibniz Computing Center (LRZ)
Munich. ER acknowledges support from the University of California
start-up funds.
\end{acknowledgments}

\appendix

\section{Reaction Coordinate Formal Solution}
\label{sec:appendix}
In this appendix we describe the formal solution of the reaction coordinate $Q\left(t\right)$ for both initial conditions (for simplicity we set $\hbar=1$). 
For the phonon Hamiltonian 
\begin{equation}
	H_{ph}=\sum_{j}\omega_{j}\left(b_{j}^{\dagger}b_{j}+\frac{1}{2}\right)+d^{\dagger}d\sum_{j}M_{j}\left(b_{j}^{\dagger}+b_{j}\right),
\end{equation}
the equation of motion for $b_{j}\left(t\right)$ reads 
\begin{equation}
	\dot{b}_{j}\left(t\right)=-i\omega_{j}b_{j}-iM_{j}d^{\dagger}d
\end{equation}
with the formal solution
\begin{equation}
	\label{eq:formal-sol}
	b_{j}\left(t\right)=b_{j}\left(0\right)e^{-i\omega_{j}t}-iM_{j}\int_{0}^{t}n_{d}\left(\tau\right)e^{-i\omega_{j}\left(t-\tau\right)}\mbox{d}\tau.
\end{equation} 
We will define the dimensionless average position of the phonon mode $j$ with   $\left\langle x_{j}\left(t\right)\right\rangle =\frac{\left\langle b_{j}^{\dagger}\left(t\right)+b_{j}\left(t\right)\right\rangle }{\sqrt{2}}$
and the corresponding reaction coordinate $\left\langle Q\left(t\right)\right\rangle =\frac{\sum_{j}M_{j}\left\langle x_{j}\right\rangle }{\sqrt{\sum_{j}M_{j}^{2}}}$
such that:
\begin{equation}
	\left\langle Q\left(t\right)\right\rangle =\frac{\Re\left\{ \sum_{j}M_{j}\left\langle b_{j}\left(t\right)\right\rangle \right\} }{\sqrt{2\sum_{j}M_{j}^{2}}}.
\end{equation}
Using Eq.~(\ref{eq:formal-sol}) we find

\begin{eqnarray}
	\left\langle Q\left(t\right)\right\rangle  & = & \frac{\sqrt{2}}{\sqrt{\sum_{j}M_{j}^{2}}}\sum_{j}M_{j}\cos\omega_{j}t\left\langle b_{j}\left(0\right)\right\rangle \nonumber \\
	& - & \frac{\sqrt{2}}{\sqrt{\sum_{j}M_{j}^{2}}}\int_{0}^{t}\left\langle n_{d}\left(t\right)\right\rangle \sum_{j}M_{j}^{2}\sin\omega_{j}\left(t-\tau\right)\mbox{d}\tau.\nonumber \\
\end{eqnarray}
The equation above can be re-written in terms of the bath autocorrelation function given in Eq.~(\ref{eq:alpha-t}) using the relations:
\begin{eqnarray*}
	\sum_{j}M_{j}^{2}e^{-i\omega_{j}t} & = & \frac{1}{\pi}\int_{0}^{\infty}\pi\sum_{\alpha}M_{j}^{2}\delta\left(\omega-\omega_{j}\right)e^{-i\omega t}\mbox{d}\omega\\
	& = & \frac{1}{\pi}\int_{0}^{\infty}J\left(\omega\right)e^{-i\omega t}\mbox{d}\omega\\
	& = & \alpha\left(t\right)
\end{eqnarray*}
and 
\begin{equation}
	\sum_{j}\frac{M_{j}^{2}}{\omega_{j}}e^{-i\omega_{j}t}=  -i\int_{0}^{t}\mbox{d}\tau\alpha\left(\tau\right)+\lambda,
\end{equation}
to reduce the formal solution of $\left\langle Q\left(t\right)\right\rangle$ into
\begin{eqnarray}
	Q\left(t\right) & = & \sqrt{\frac{2}{\alpha\left(0\right)}}\Im\int_{0}^{t}\delta n\left(\tau\right)\alpha\left(t-\tau\right)\mbox{d}\tau\nonumber \\
	& + & \sqrt{\frac{2}{\alpha\left(0\right)}}n_{\infty}\mbox{\ensuremath{\Im\int_{0}^{t}\alpha\left(\tau\right)}\mbox{d}\ensuremath{\tau}}
\end{eqnarray}

The equation above corresponds to  the non-shifted bath initial condition \emph{i.e.} $\left\langle b_{j}\left(0\right)\right\rangle =0$, while for the shifted-bath $\left(\left\langle b_{j}\left(0\right)\right\rangle =-\frac{M_{j}}{\omega_{j}}\right),$ the solution reads:
\begin{eqnarray}
	\left\langle Q\left(t\right)\right\rangle  & = & -\sqrt{\frac{2}{\alpha\left(0\right)}}\lambda+\sqrt{\frac{2}{\alpha\left(0\right)}}\Re\left\{ i\int_{0}^{t}\mbox{d}\tau\alpha\left(\tau\right)\right\} \nonumber \\
	& + & \sqrt{\frac{2}{\alpha\left(0\right)}}\Im\int_{0}^{t}\delta n\left(\tau\right)\alpha\left(t-\tau\right)\mbox{d}\tau\nonumber \\
	& + & \sqrt{\frac{2}{\alpha\left(0\right)}}n_{\infty}\mbox{\ensuremath{\Im\int_{0}^{t}\alpha\left(\tau\right)}\mbox{d}\ensuremath{\tau}}.
\end{eqnarray}
In the above $\delta n\left(t\right)=\left\langle n_{d}\left(t\right)\right\rangle -n_{\infty}$
with $n_{\infty}=\lim_{t\rightarrow\infty}\left\langle n_{d}\left(t\right)\right\rangle $
and $\Re\left\{ ...\right\} \left(\Im\left\{ ...\right\} \right)$
are the real and imaginary parts respectively.\\
One can look at the steady state limit using the fact that $\lim_{t\rightarrow\infty}\sqrt{\frac{2}{\alpha\left(0\right)}}\Im\int_{0}^{t}\delta n\left(\tau\right)\alpha\left(t-\tau\right)\mbox{d}\tau\rightarrow0$
since $\delta n(\tau)$ and $\alpha\left(\tau\right)$ approach zero for $\tau \to \infty$ and 
\begin{equation}
	\int_{0}^{\infty}\alpha\left(\tau\right)\mbox{d}\tau=-i\lambda
\end{equation}
to obtain 
\begin{eqnarray}
	\Delta Q & = & -\sqrt{\frac{2}{s\lambda\omega_{c}}}\lambda\Delta n\\
	& = & -\sqrt{\frac{2\lambda}{s\omega_{c}}}\Delta n
\end{eqnarray}
where $\Delta Q$ and $\Delta n_{d}$ are the steady state difference between the shifted and the non-shifted phonon bath for the reaction coordinate and the dot population respectively.

\bibliographystyle{apsrev}
\bibliography{diff-s}

\end{document}